\documentclass[preprint,showpacs,preprintnumbers,amsmath,amssymb]{revtex4}

\usepackage{graphicx}% Include figure files
\usepackage{dcolumn}% Align table columns on decimal point
\usepackage{bm}% bold math

\begin{document}

\preprint{APS/123-QED}

\title{ Anisotropic distribution of nucleon participating in elliptical flow\\}%Force line breaks with\\

\author{Anupriya Jain}
\author{Suneel Kumar}%
 \email{suneel.kumar@thapar.edu}
\affiliation{%
School of Physics and Material Science, Thapar University, Patiala-147004, Punjab (India)\\
}%

%\author{Rajeev K. Puri}
%\affiliation{
%Department of Physics, Panjab University, Chandigarh-160014, India\\
% with \\
%}%
\date{\today}% It is always \today, today,

\begin{abstract}
Using the isospin dependent quantum molecular dynamics model, we study the effect of charge asymmetry and isospin dependent cross-section on $\frac{dN}{d(\langle Cos2\phi \rangle)}$ and $\frac{dN}{p_tdp_t}$. Simulations have been carried out for the reactions of $^{124}X_{m}+^{124}X_{m}$, where m = (47, 50 and 59) and $^{40}S_{16}+^{40}S_{16}$. Our study shows that these parameters depend strongly on the isospin of cross-section and charge asymmetry. The distribution of nucleons and fragments is not symmetric around the beam axis. 
\end{abstract}

\pacs{25.70.-z, 25.70.Pq, 21.65.Ef}% PACS, the Physics and Astronomy
\maketitle
\baselineskip=1.5\baselineskip\
\section{Introduction}

It is well known that collective flow is an important observable in heavy ion collisions (HIC) and it can give some essential information
about the nuclear matter, such as the nuclear equation of state \cite{persram, lukasik, yan, chen}. For the last few years, collective flowhas been used as a powerful tool to explore the nuclear equation of state (EOS) as well as in medium nucleon-nucleon cross-section \cite{vermani} Anisotropic flow is defined as the different $n^{th}$ harmonic coefficient $v_{n}$ of the Fourier expansion for the particle invariant azimuthal distribution \cite{voloshin}:

\begin{equation}
\frac{dN}{d\phi} = 1+2 \sum_{n=1}^{\infty}v_n Cos(n\phi)
\end{equation}
where $\phi$ is the azimuthal angle between the transverse momentum of the particle and the reaction plane. Note that the z-axis is defined as the direction along the beam and the impact parameter axis is labelled as x-axis. Anisotropic flows generally depend on both particle transverse momentum and rapidity, and for a given rapidity the anisotropic flows at transverse momentum $p_{t}$
($p_{t}$=$\sqrt{(p_x^2 + p_y^2)}$, where $p_{x}$ and $p_{y}$ are projections of particle transverse momentum in and perpendicular to the reaction plane, respectively. The first harmonic coefficient $v_{1}$ is called directed flow parameter. Directed flow is the measure of the collective motion of the particles in the reaction plane. This flow is reported to diminish at higher incident energies due to the large beam rapidity \cite{sood}. The second harmonic coefficient $v_{2}$ is called the elliptic flow parameter $v_{2}$. Elliptic flow in heavy ion collisions is a measure of the azimuthal angular anisotropy of particle distribution in momentum space with respect to the reaction plane \cite{zheng}. The elliptic flow at intermediate energy HIC is complex phenomenon because it is determined by the interplay among fireball expansion, collective rotation, the shadowing of spectators, Coulomb repulsion, and so on. Both the mean field and two-body collision parts play important roles: the mean field plays a dominant role at low energies, and then gradually the two-body collisions become dominant with energy increase. The transverse radial dependent transverse velocity can reflect the correlation between spacial and momentum coordinates, and reveal the force change on fragments along the transverse radius. The magnitude of the elliptic flow depends on both initial spatial asymmetry in non-central collisions and the subsequent collective interactions. Experimentally observed out-of-plane emission termed as squeeze-out was observed by SATURNE (France) by DIOGENE collaboration \cite{gosset}. The Plastic-Ball group at the BEVELAC in Berkley were the first one to quantify the squeeze-out in symmetric systems \cite{gutbrod}. The elliptic flow is sensitive to the properties of the dense matter formed during the initial stage of heavy ion collision \cite{kolb} and parton dynamics \cite{zhang} at Relativistic Heavy Ion Collider (RHIC) energies.\\
C. Pinkenburg {\it et al.,} \cite{Pinkenburg} have measured an elliptic flow excitation function for midcentral collisions of Au + Au at 2, 4, 6, and 8 GeV/nucleon respectively. The excitation function exhibits a transition from negative to positive elliptic flow with transition energy $E_{trans}$ = 4A GeV.\\
J. Lukasik {\it et al.,} \cite{lukasik} studied the distributions of the squeeze angle for incident energies from 40 to 150 MeV/nucleon, their study revealed that, the minima at $\phi$ = $\pi$/2, occur at lower energies and more peripheral impact parameters while peaks at $\phi$ = $\pi$/2, most strongly pronounced in the more central bins at the higher incident energies.\\
J. H. Chen {\it et al.,} \cite{chen} studied the azimuthal angular distribution of raw $\phi$ yields with respect to the event plane, they showed that, the finite resolution in the approximation of event plane as reaction plane smears out the azimuthal angular distribution and leads to a lower value in the apparent anisotropy parameters. In this paper our aim is to check that, whether the distribution of nucleon contributing to elliptical flow are distributed equally along the ellipse or not. Our study is performed within the framework of IQMD \cite{hartnack} model which is the improved version of QMD \cite{aichelin} model.\\
%%%%%%%%%%%%%%%%%%%%%%%%%%%%%%%%%%%%%%%%%%%%%%%%%%%%%%%%%%%%%%%%%%%%%%%%%
\section{Results and Discussion}
For the present analysis, simulations are carried out for the reactions of $^{124}X_{m}+^{124}X_{m}$, where m = (47, 50 and 59) and $^{40}S_{16}+^{40}S_{16}$. The phase space generated by the IQMD model has been analyzed using the minimum spanning tree (MST) \cite{rkpuri} method. 
The elliptical flow is defined as the average difference between the square of x and y components of the particles transverse momentum. Mathematically, it can be written as \cite{vermani}:

\begin{equation}
\langle v_2 \rangle = <Cos2\phi> = \langle \frac{p_x^2 - p_y^2}{p_x^2 + p_y^2}\rangle
\end{equation}

where $p_{x}$ and $p_{y}$ are the x and y components of the momentum. The positive value of elliptical flow describes the eccentricity of an ellipse-like distribution and indicates in-plane enhancement of the particle emission. On the other hand, a negative value of $v_{2}$ shows the squeeze-out effects perpendicular to the reaction plane. Obviously, zero value corresponds to an isotropic distribution. 
%%%%%%%%%%%%%%%%%%%%%%%%%%%%%%%%%%%%%%%%%%%%%%%%%%%%%%%%%%%%%%%%%%%%%%%%%
\begin{figure}
\hspace{-2.0cm}\includegraphics[scale=0.45]{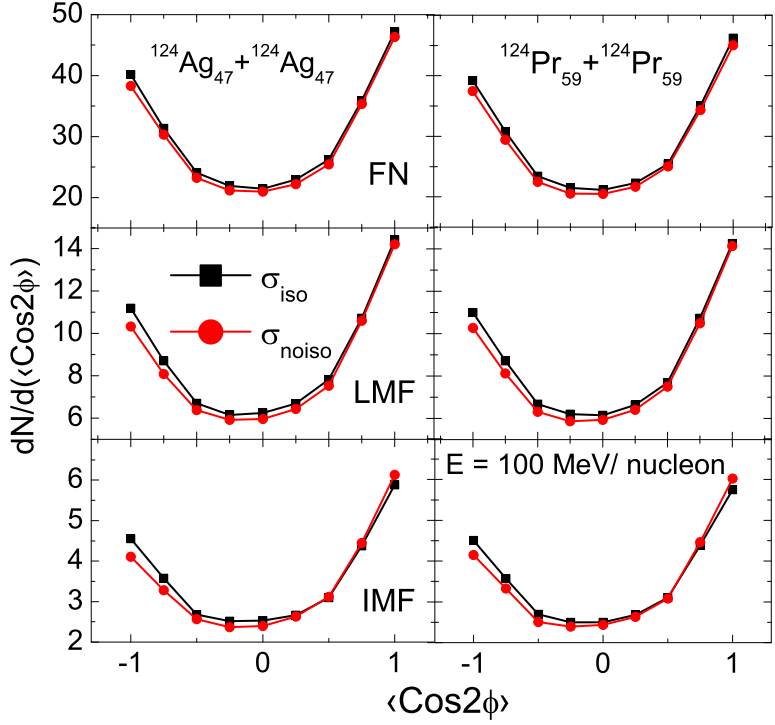}
\caption{\label{Fig:1} (color online) Azimuthal angle dependence of $\frac{dN}{d(\langle Cos2\phi \rangle)}$, for free nucleons (upper panel), LMF's (middle) and IMF's (lower panel).}
\end{figure}

To study the effect of isospin dependent cross-section and charge asymmetry on $\frac{dN}{d(\langle Cos2\phi \rangle)}$, we display in fig.1, the azimuthal angle dependence of $\frac{dN}{d(\langle Cos2\phi \rangle)}$, for free nucleons (A = 1) (upper panel), LMF's (2 $\leq$ A $\leq$ 4)(middle) and IMF's (5 $\leq$ A $\leq$ $A_{tot}$/6) (lower panel) at an incident energy E = 100 MeV/nucleon for the reactions of $^{124}Ag_{47}+^{124}Ag_{47}$ (left panels) and $^{124}Pr_{59}+^{124}Pr_{59}$ (right panels). Figure reveal:\\
(a) Minima at 2$\phi$=$\pi$/2 indicate predominantly in-plane emission, while the peaks at 2$\phi$= 0 and $\pi$, corresponds to a preference for azimuthal emission in-plane and perpendicular to the reaction plane, the so-called squeeze-out.\\
(b) Peak is more pronounced at 2$\phi$= 0 than at 2$\phi$= $\pi$, which indicates that number of particles emmited in-plane are large as compare to the number of particles emitted out-of-plane. This means that ellipse formed is not symmetric around the Z-axis i.e in the collision of symmetric nuclei the distribution of nucleon after the collision in momentum space is not uniform.\\
(c) There is a very little influence of charge asymmetry on the variation of $\frac{dN}{d(\langle Cos2\phi \rangle)}$ with $\langle Cos2\phi \rangle$. $\frac{dN}{d(\langle Cos2\phi \rangle)}$ is more for neutron rich system $^{124}Ag_{47}+^{124}Ag_{47}$ than neutron deficient system $^{124}Pr_{59}+^{124}Pr_{59}$ due to increase in repulsive forces.\\
(d) $\frac{dN}{d(\langle Cos2\phi \rangle)}$ is sensitive to different nucleon-nucleon cross-sections. Its value is more in case of isospin dependent cross-section. This happens because in the case of isospin dependent cross-section, neutron-proton cross-section is three times larger compared to neutron-neutron and proton-proton cross-section that will enhance binary collisions. Moreover, in case of neutron rich system, due to more repulsion more squeeze-out can be seen.\\
%%%%%%%%%%%%%%%%%%%%%%%%%%%%%%%%%%%%%%%%%%%%%%%%%%%%%%%%%%%%%%%%%%%%%%%
\begin{figure}
\hspace{-2.0cm}\includegraphics[scale=0.45]{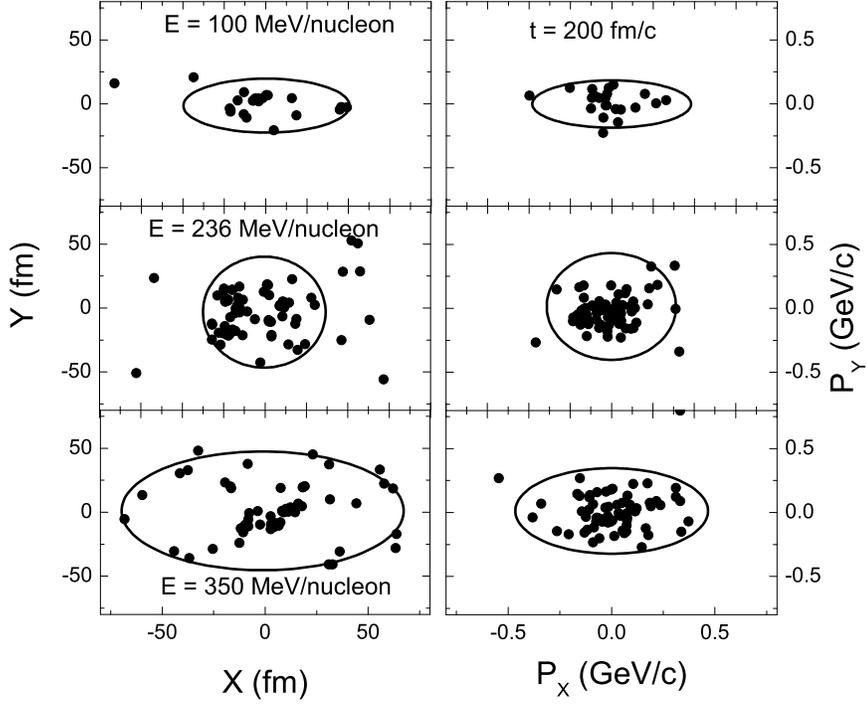}
\caption{\label{Fig:2} Phase space distribution of nucleons in X-Y plane(left panel) and $P_{X}$-$P_{Y}$ plane (right panel). The reaction under study is $^{124}Ag_{47}+^{124}Ag_{47}$. The panels from top to bottom are representing the phase space of nucleons at different energies.}
\end{figure}
  
To further strengthen our interpretation of the results of fig.1, we display in fig.2, the final phase space of nucleons for X-Y plane and $P_{X}$-$P_{Y}$ plane for the reaction of $^{124}Ag_{47}+^{124}Ag_{47}$ at an incident energy of 100 MeV/nucleon (below transition energy), 236 MeV/nucleon (at transition energy) and 350 MeV/nucleon (above transition energy) in upper, middle and below panels respectively for a randomly selected event. It has been observed that the distribution of the nucleons is ellipse like for E = 100 and 350 MeV/nucleon i.e below and above the transition energy and spherical for E = 236 MeV/nucleon i.e at the transition energy. 
%%%%%%%%%%%%%%%%%%%%%%%%%%%%%%%%%%%%%%%%%%%%%%%%%%%%%%%%%%%%%%%%%%%%%%%
 \begin{figure}
\hspace{-2.0cm}\includegraphics[scale=0.45]{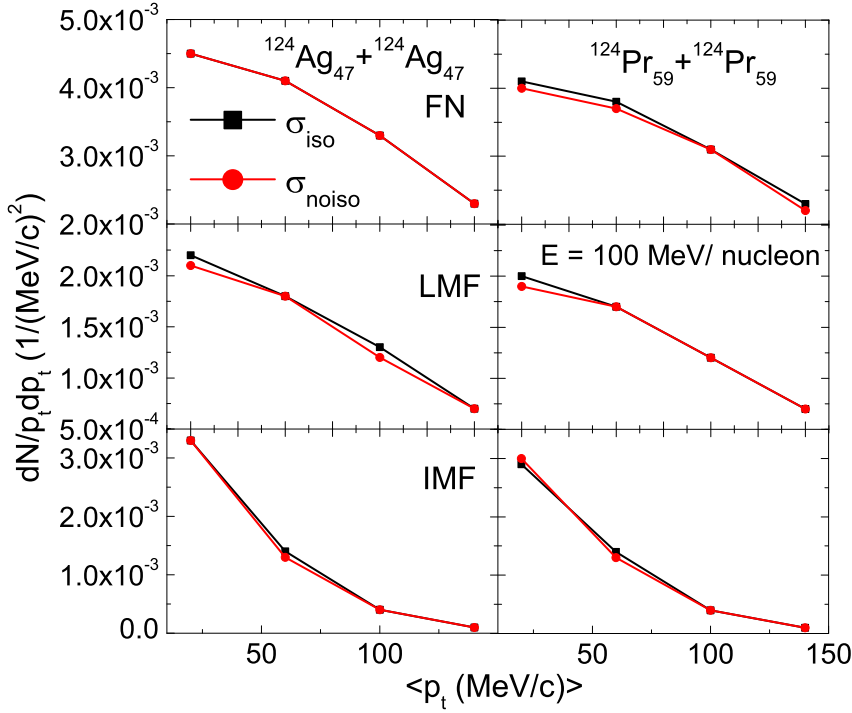}
\caption{\label{Fig:3} (color online) Transverse momentum dependence of $\frac{dN}{p_{t}dp_{t}}$ for the reactions of $^{124}Ag_{47}+^{124}Ag_{47}$ (left) and $^{124}Pr_{59}+^{124}Pr_{59}$ (right).}
\end{figure}

To study the effect of isospin dependence of cross-section and charge asymmetry on $\frac{dN}{p_{t}dp_{t}}$, we display in Fig.3, the transverse momentum dependence of $\frac{dN}{p_{t}dp_{t}}$ for the reactions of $^{124}Ag_{47}+^{124}Ag_{47}$ and $^{124}Pr_{59}+^{124}Pr_{59}$ at an incident energy E = 100 MeV/nucleon. The figure reveal following points:\\
(a) As the transverse momentum increases $\frac{dN}{p_{t}dp_{t}}$ decrease. Which is quite obvious, because with increase in transverse momentum, the number of particles in that particular bin with large transverse momentum decreases. This shows that after the collision, momentum is not equally transfered among the nucleon. Some nucleon suffer hard collision while other suffer soft collision.\\
(b) The value of $\frac{dN}{p_{t}dp_{t}}$ is more for neutron rich system
 $^{124}Ag_{47}+^{124}Ag_{47}$ than neutron deficient system $^{124}Pr_{59}+^{124}Pr_{59}$ due to increase in repulsive forces among nucleon.\\
(c) $\frac{dN}{p_{t}dp_{t}}$ is sensitive to different nucleon-nucleon cross-section. Its value is more in case of isospin dependent cross-section. This happens because in the case of isospin dependent cross-section, neutron-proton cross-section is three times large compared to neutron-neutron and proton-proton cross-section that will enhance binary collisions. But this increase is not uniform for free nucleons.\\
%%%%%%%%%%%%%%%%%%%%%%%%%%%%%%%%%%%%%%%%%%%%%%%%%%%%%%%%%%%%%%%%

\begin{figure}
\hspace{-2.0cm}\includegraphics[scale=0.45]{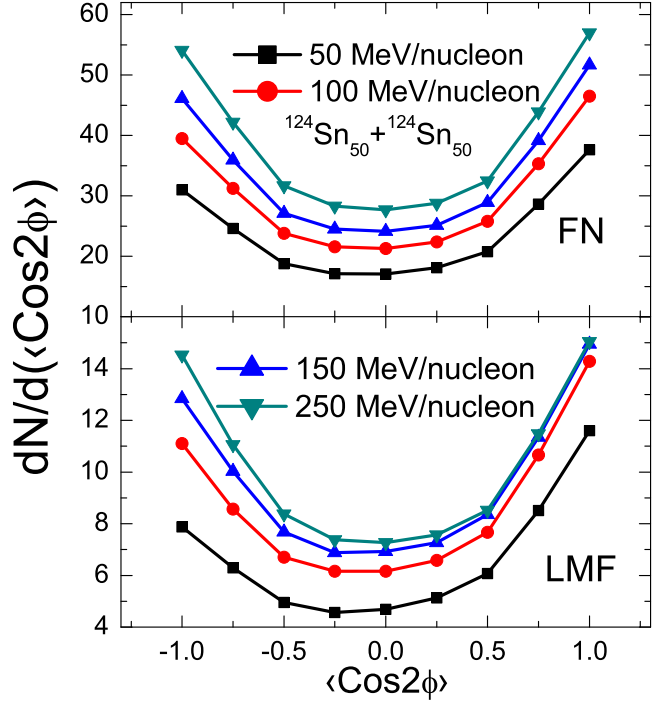}
\caption{\label{Fig:4} (color online) Azimuthal angle dependence of $\frac{dN}{d(\langle Cos2\phi \rangle)}$ for free nucleons (upper panel) and LMF's (lower panel) at incident energies 50, 100, 150 and 250 MeV/nucleon.}
\end{figure}

In fig.4, we display the azimuthal angle dependence of $\frac{dN}{d(\langle Cos2\phi \rangle)}$ for free nucleons and LMF's at incident energies 50, 100, 150 and 250 MeV/nucleon for the reaction of $^{124}Sn_{50}+^{124}Sn_{50}$. We note:\\
 As the energy increases, slope of the curve increases for both free nucleons and LMF's. This happens because, as the energy increases the thrust will also increases which will enhance the out-of-plane flow of the nucleon. 
The increase in slope is uniform in case of free nucleons but non-uniform in case of LMF's. Which indicates that emission of free nucleon is symmetrical but emission of LMF's is not symmetrical about reaction plane. This is something interesting which we were not expecting.\\
%%%%%%%%%%%%%%%%%%%%%%%%%%%%%%%%%%%%%%%%%%%%%%%%%%%%%%%%%%%%%%%%
\begin{figure}
\hspace{-2.0cm}\includegraphics[scale=0.45]{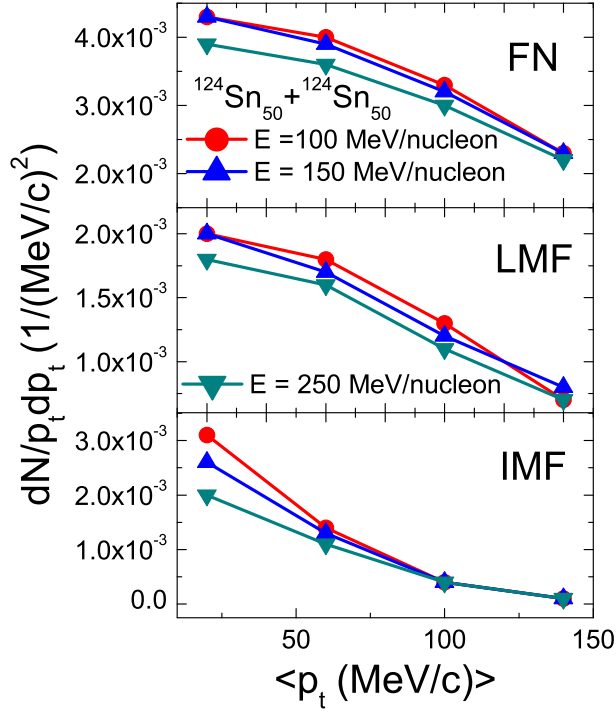}
\caption{\label{Fig:5} (color online) Transverse momentum dependence of $\frac{dN}{p_{t}dp_{t}}$ for free nucleons, LMF's and IMF's at incident energies E = 100, 150 and 250 MeV/nucleon.} 
\end{figure}
In fig.5, we display the transverse momentum dependence of $\frac{dN}{p_{t}dp_{t}}$ for free nucleons, LMF's and IMF's at incident energies E = 100, 150 and 250 MeV/nucleon for the reaction of $^{124}Sn_{50}+^{124}Sn_{50}$. We note that as the energy increases, $\frac{dN}{p_{t}dp_{t}}$ decreases. This happens because, with increase in energy the particles with large transverse momentum will decrease in that particular bin thus the curve shift downward for free nucleons, LMF's and IMF's. 
%%%%%%%%%%%%%%%%%%%%%%%%%%%%%%%%%%%%%%%%%%%%%%%%%%%%%%%%%%%%%%%%%%%%%%%%%
\begin{figure}
\hspace{-2.0cm}\includegraphics[scale=0.45]{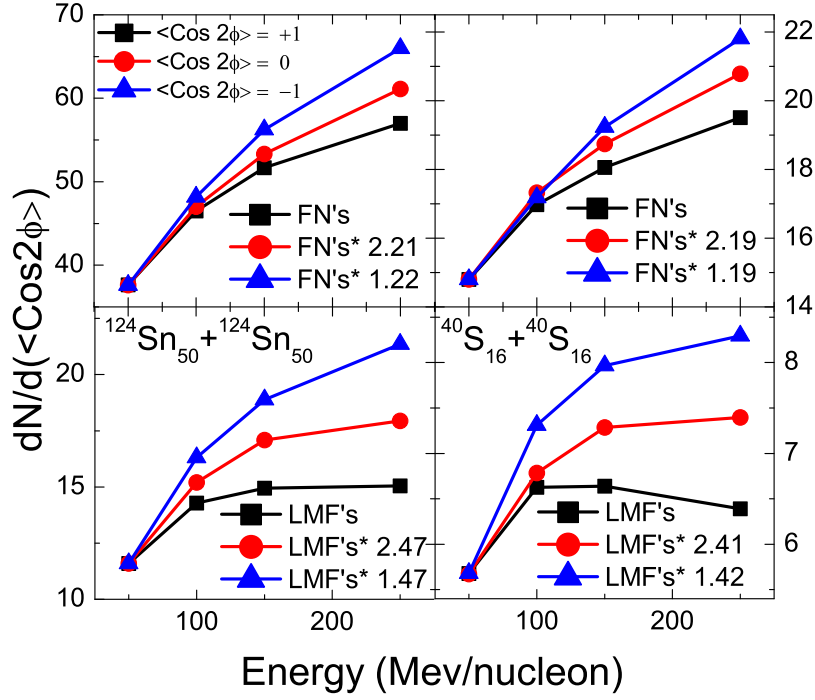}
\caption{\label{Fig:6} (color online) The energy dependence of $\frac{dN}{d(\langle Cos2\phi \rangle)}$ for the reactions of  $^{124}Sn_{50}+^{124}Sn_{50}$ and $^{40}S_{16}+^{40}S_{16}$.}
\end{figure}

To further strengthen our interpretation of the results, we display in fig.6 the energy dependence of $\frac{dN}{d(\langle Cos2\phi \rangle)}$ for free nucleons and LMF's for the reactions of $^{124}Sn_{50}+^{124}Sn_{50}$ (N/Z = 1.48) and $^{40}S_{16}+^{40}S_{16}$ (N/Z = 1.5) for $\langle Cos2\phi \rangle$ = -1, 0, +1. One can note that, once the free nucleons and LMF's at $\langle Cos2\phi \rangle$ = 0 and -1 are normalized with free nucleons and LMF's at $\langle Cos2\phi \rangle$ = +1 at the starting point of energy, we see that their behavior with respect to the energy is similar for both the free nucleons and LMF's. Although, the N/Z of both the reactions are nearly equal but the behavior of energy dependence of $\frac{dN}{d(\langle Cos2\phi \rangle)}$ is different for both the reactions. One can see from the fig.5, that out-of-plane emission is more in case of $^{124}Sn_{50}+^{124}Sn_{50}$ than in-plane emission. But the behavior is entirely different for  $^{40}S_{16}+^{40}S_{16}$ where the in-plane emission is more than out-of-plane emission. The distribution of free nucleon is equal for in-plane and out-of-plane flow with respect to distribution corresponding to vanishing flow. But for LMF's the distribution is asymmetric. \\  
%%%%%%%%%%%%%%%%%%%%%%%%%%%%%%%%%%%%%%%%%%%%%%%%%%%%%%%%%%%%%%%%%%%%%%%%%
\section{Summary}

Using the isospin dependent quantum molecular dynamics model, we have studied the effect of charge asymmetry and isospin dependent cross-section on $\frac{dN}{d(\langle Cos2\phi \rangle)}$ and $\frac{dN}{p_tdp_t}$. Simulations have been carried out for the reactions of $^{124}X_{m}+^{124}X_{m}$, where m = (47, 50 and 59) and $^{40}S_{16}+^{40}S_{16}$. Our study showed that distribution of nucleons and fragments is not symmetric in space for in-plane and out-of-plane emission.\\
%%%%%%%%%%%%%%%%%%%%%%%%%%%%%%%%%%%%%%%%%%%%%%%%%%%%%%%%%%%%%%%%%%%%%%%%%
{\large{\bf Acknowledgment}}

This work has been supported by a grant from the university grant commission (UGC), Government of India [Grant No. 39-858/2010(SR)].\\

\noindent

%%%%%%%%%%%%%%%%%%%%%%%%%%%%%%%%%%%%%%%%%%%%%%%%%%%%%%%%%%%%%%%%%%%%%%%%%%

\end{document}